\documentclass[journal,comsoc]{IEEEtran}
\usepackage{algorithm,algorithmicx,amsbsy,amsmath,amssymb,epsfig,bbm,mathrsfs,multirow,amsthm,cite,}

\usepackage{subcaption}
\usepackage[labelformat=simple]{subcaption}

\usepackage{hyperref}
\usepackage{eurosym}
\hypersetup{
	colorlinks=true,
	linkcolor=black,
	filecolor=black,   
	urlcolor=black,citecolor=black,}

\newtheorem{proposition}{Proposition}
\DeclareMathAlphabet\mathbfcal{OMS}{cmsy}{b}{n}
\usepackage{array}

\begin{document}
	
	\title{Jointly Optimize Coding and Node Selection for Distributed Computing over Wireless Edge Networks}
	
	\author{\IEEEauthorblockN{Cong T. Nguyen, Diep N. Nguyen, Dinh Thai Hoang, Hoang-Anh Pham and Eryk Dutkiewicz \vspace*{-4mm}}
	\thanks{Cong T. Nguyen, Diep N. Nguyen, Dinh Thai Hoang, and Eryk Dutkiewicz are with University of Technology Sydney, Australia (email: cong.nguyen@student.uts.edu.au, Hoang.Dinh, Diep.Nguyen, and Eryk.Dutkiewicz@uts.edu.au). Cong T. Nguyen and Hoang-Anh Pham is with Ho Chi Minh City University of Technology, Vietnam (email: ntcong.sdh19@hcmut.edu.vn and anhpham@hcmut.edu.vn).}
}

	\maketitle
	
	\begin{abstract}
This work aims to jointly optimize the coding and node selection to minimize the processing time for distributed computing tasks over wireless edge networks. Since the joint optimization problem formulation is NP-hard and nonlinear, we leverage the discrete characteristic of its decision variables to transform the problem into an equivalent linear formulation. This linearization can guarantee to find the optimal solutions and significantly reduce the problem’s complexity. Simulations based on real-world datasets show that the proposed approach can reduce the total processing time up to 2.3 times compared with that of state-of-the-art approach. 	
\end{abstract}
	
	
	\section{Introduction}
	\label{sec:introduction}
	With the ability to utilize multiple wireless devices (e.g., edge and mobile devices) to simultaneously execute intensive computing tasks, distributed computing has recently become a highly-effective approach for large-scale computations in wireless edge networks. Compared to centralized computing methods, distributed computing is more fault-tolerant, i.e., it can still function even when several nodes fail. Moreover, distributed computing systems are more scalable, i.e., more devices can be easily added~\cite{Cdcsurvey}. With these outstanding advantages, distributed computing has been widely applied in areas such as Internet-of-Things~\cite{iot}, edge intelligent~\cite{edge}, and especially with recent advances in distributed learning~\cite{ML}. 
	
	However, the uncertainties of computing processes (e.g., inconsistent computation time and/or failures) and wireless connections in wireless edge networks lead to less predictable latency and cause serious straggling problems. Specifically, straggling edge nodes with unstable wireless connections and/or computing resources can dramatically slow down computing processes for the whole network. To address this issue, coded distributed computing~\cite{Cdcsurvey} (CDC) has recently emerged to be a promising solution. The core idea of CDC is to use advanced coding theoretic techniques to assign redundant workload to edge nodes to compensate for the stochastic computation and transmissions, thereby improving the latency and stability of conventional distributed computing~\cite{Cdcsurvey}. Among these techniques, the maximum distance separable (MDS) code has been widely adopted in the literature~\cite{Cdcsurvey}. By using the MDS code, a task can be encoded into sub-tasks and distributed to $n$ nodes, and the final result can be obtained by decoding \emph{any} $k$ first results from the $n$ sub-tasks ($k \leq n$). 
	
	In the MDS, choosing values of the $(n,k)$ code has a significant impact on the total processing time of the CDC process. Particularly, since the size of each sub-task is inversely proportional to $k$~\cite{Cdcsurvey}, choosing a high value of $k$ means that each node needs to solve a smaller sub-task, thereby reducing the computation time at each node. However, in this case, the master node needs to wait for more nodes to send back their computing results. This can potentially lead to a longer delay if there are many slow computing nodes and unreliable communication links in the system. Moreover, most of existing approaches, e.g.,~\cite{r1,r3,r4}, only focus on optimizing the $(n,k)$ code without considering node selection, i.e., they assume that the nodes are identical in terms of computing resource and communication links. However, in practice, edge nodes have dissimilar hardware configurations and communication links. Consequently, this significantly hinders the applicability of CDC in heterogeneous environments such as wireless edge networks. Therefore, node selection and the $(n,k)$ coding need to be jointly considered and optimized, which is a very challenging task. To the best of our knowledge, our paper is the first work in the literature aiming to optimize both the coding and the node selection for CDC-based wireless edge networks.
	
	In this letter, we develop a highly-effective framework to jointly optimize the $(n,k)$ code and node selection for a CDC-based wireless edge network. In particular, we first formulate the joint code and node selection optimization problem as a Mixed Integer Non-linear Programming (MINLP). Since this problem is NP-hard and nonlinear, we leverage the discrete characteristic of its decision variables to develop a linearization approach, thereby transforming the MINLP into an equivalent Mixed-Integer Linear Programming (MILP) problem. This can significantly reduce the complexity of the original problem, and it can be efficiently solved by commercial MILP solvers to obtain the optimal solutions. Simulations based on real-world datasets are also conducted to evaluate and compare the performance of our proposed approach with other existing approaches. The results show that our proposed approach can outperform state-of-the-art approach by up to 2.3 times.
	
	\begin{figure}[!]
		\includegraphics[width=.45\textwidth]{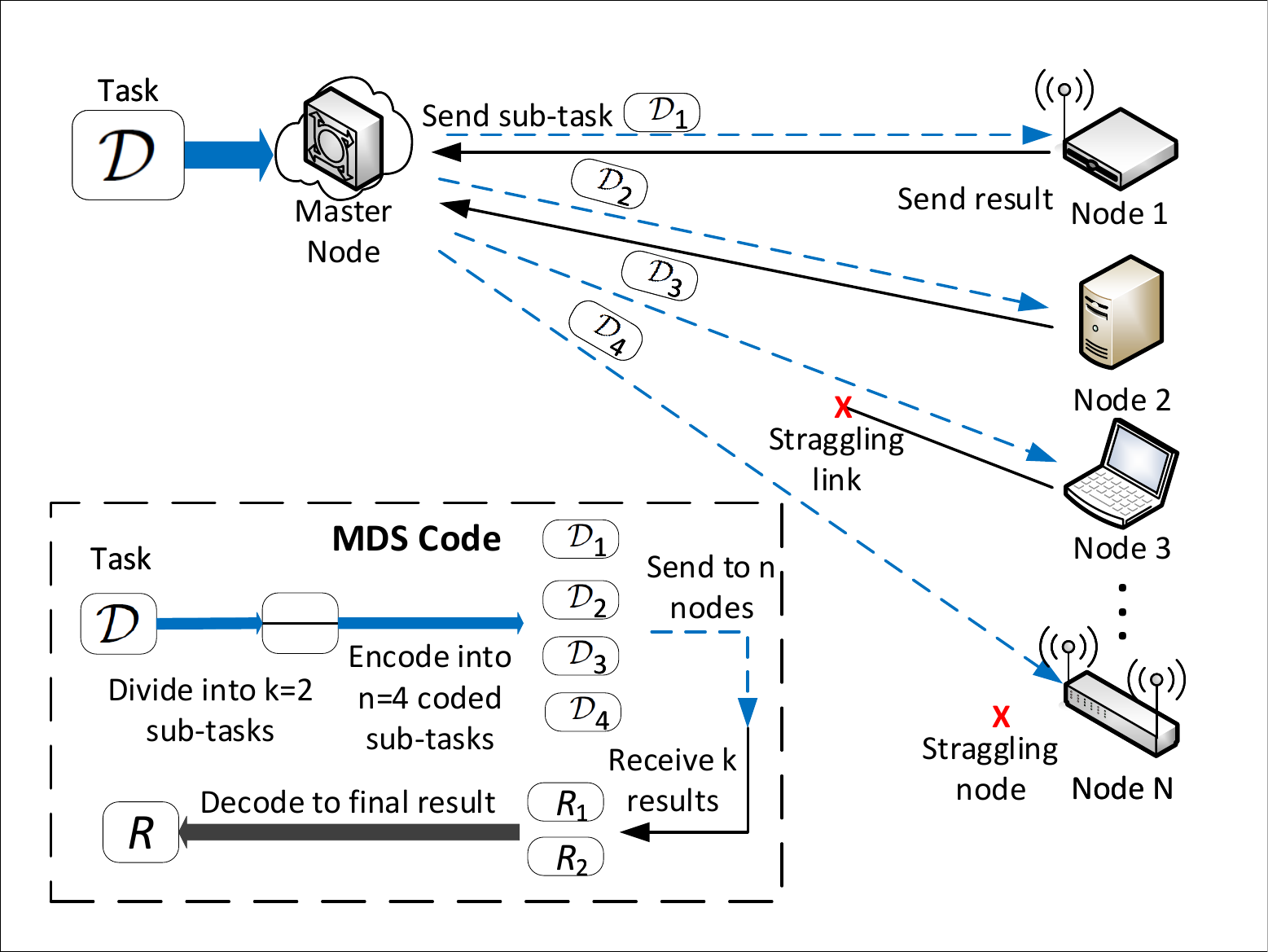}
		\centering
		\caption{Illustration of a coded distributed computing system.}
		\label{Fig:system}
	\end{figure}

	\section{System Model}
	\label{sec:system model}

	\subsection{System Overview}
 	We consider a CDC-based system over a wireless edge network consisting of a server and a set $\mathcal{N}$ of $N$ edge nodes. Upon receiving a computing task $\mathcal{D}$, the server first divides it into $k$ sub-tasks using the $(n,k)$ MDS code. Then, these $k$ sub-tasks are encoded into $n$ identical coded sub-tasks and sent to $n$ nodes ($n\leq N$). With the MDS code, the size of the sub-task sent to node $i$ is $d=D/k$, where $D$ is the size of task $\mathcal{D}$. The total time $t_i$ needed for node $i$ to complete a sub-task is $t_i= t^s_i+t^c_i$, 
 where $t^s_i$ is the total communication time it takes for the server to send the sub-task to node $i$ and for node $i$ to send the result back to the server. Moreover, $t^c_i$ is the computation time for node $i$ to obtain the result for its assigned sub-task. After receiving the assigned sub-tasks, the nodes then perform the computations locally and send the results to the server once they finish. The server only needs the first $k$ ($k \leq n$) results among those of the $n$ sub-tasks to be able to derive the final result of $\mathcal{D}$. Let $\mathcal{K}$ denote the set of the first $k$ nodes that successfully send the result back to the server. Then, the total processing time $T$ can be defined by
 $T=\max \{t_k: \forall k\in \mathcal {K}\}$. For example, if we have $(n,k)=(4,2)$ and $\{t_i\}=\{1,5,9,3\}$, then $T=t_4=3$. 
 \subsection{Computation Model}
 The computation time of a node consists of the deterministic and stochastic computation time, i.e., $t^c_i= t^d_i+t^r_i$.
 The deterministic computation time is given by $t^d_i=d/\eta_i=D/k\eta_i$,
 where $\eta_i$ is the number of computations that node $i$ can perform per second, and $D$ is the size of task $\mathcal{D}$. The stochastic computation time is assumed to follow an exponential distribution, i.e., $p_{t^r_i}=\lambda_i e^{-\lambda_i t}$, where $\lambda_i=\alpha_i\eta_i/d$. $\alpha_i$ represents the stochastic component of computation time coming from random memory access~\cite{jsac}. Thus, the expected value of $t^r_i$ is 	$\mathrm{E}[t^r_i]=1/\lambda_i=D/k\eta_i\alpha_i$.
	\subsection{Communication Model}
	The communication time between a node and the server depends on the communication link between them, which might be different for each node. Typically, the communication time can be expressed as $t^s_i= \tau_i (N^d_i + N^u_i)$,
	where $\tau_i$ is the deterministic time to upload/download a sub-task. $N^d_i$ and $N^u_i$ are the numbers of attempts required for a successful transmission. We assume that $N^d_i$ and $N^u_i$ follow a geometric distribution with probability $p_i$, and $N^d_i = N^u_i$~\cite{jsac}. Then, the expected value of $t^s_i$ is $\mathrm{E}[t^s_i]=2\tau_i/p_i$.


	\section{Problem Formulation and Solutions}
	\label{perform}
	\subsection{Problem Formulation}
	In this work, we aim to minimize the total processing time $T$ of a task by jointly optimizing the $(n,k)$ code and the node selection for the CDC-based wireless edge network where the edge nodes have different computing power $\eta_i$ and transmission time $\tau_i$. To this end, we formulate the considered problem as model $P_1$ as follows:
		\begin{align}
		\min_{n,k,\mathbf{c},\mathbf{x}}  & T,\label{c1}\\
		\textrm{s.t.}  & \sum_{i \in \mathcal{N}}c_{i}= n,  &  \label{c2}  \\
		& x_{i} \leq c_{i},  & \forall i \in \mathcal{N},\label{c3}   \\
		& \sum_{i \in \mathcal{N}}x_{i} = k,  &  \label{c4}  \\
		&t_{i} + (1-x_{i})M \geq \dfrac{2\tau_i}{p_i} +\dfrac{D}{k\eta_i(1+\alpha_i)}, \hspace{-0.5em}&\forall i \in \mathcal{N},\label{c5}\\
		&T \geq \sum_{i \in \mathcal{N}} t_{i},&\label{c6}\\
		& k \leq n, &\label{c7} \\
		& n \leq N, &\label{c8} \\
		&c_{i}, x_{i} \in \{0,1\},  & \forall i \in \mathcal{N},   \label{c9}\\
		&n,k \in \mathbb{N}.\label{c10}
	\end{align}
	In $P_1$, objective~\eqref{c1} aims to minimize the total processing time $T$ of task $\mathcal{D}$. Constraints~\eqref{c2} to~\eqref{c4} set the values for $\mathbf{c}$ and $\mathbf{x}$ which represent the nodes selection decisions. Particularly, $c_i=1$ if node $i$ is among the $n$ nodes selected, and $x_i=1$ if node $i \in \mathcal{K}$. In this way, these constraints ensure that (i) there are $n$ nodes selected, (ii) if node $i$ is selected, then $c_i=1$, (iii) $k$ nodes are selected from $n$ nodes, and (iv) there are exactly $k$ nodes with $x_i=1$.
	
	Then, constraint~\eqref{c5} determines the processing time $t_i$ of the selected $k$ nodes. $M$ is a large number to ensure that $t_i$ is bounded for only the nodes with $x_i=1$. After the $t_i, \forall i\in \mathcal{N}$, are bounded, constraint~\eqref{c6} ensures that $T$ is bounded by the highest $t_i$. Constraints~\eqref{c7} and~\eqref{c8} set the conditions for $n$ and $k$, i.e., $k \leq n \leq N$. Finally, constraints~\eqref{c9} and~\eqref{c10} set the value domains for $n$, $k$, $c_i$, and $x_i$.
	\subsection{Complexity of the MINLP Formulation}
	From the above formulation, we can observe two characteristics regarding its complexity. First, the considered optimization problem is NP-hard as proven in Proposition 1. 
	\begin{proposition}
		The optimization problem $P_1$ is NP-hard.
	\end{proposition} 
	\begin{IEEEproof}
		The considered problem can be decomposed into two sub-problems, namely MDS code optimization and node selection. The node selection sub-problem, defined by~\eqref{c3}-\eqref{c10} is equivalent to the 0-1 knapsack problem~\cite{MILP}. Therefore, the node selection sub-problem is NP-hard. Consequently, the considered joint optimization problem is NP-hard. 
	\end{IEEEproof}
	Moreover, constraint~\eqref{c5} is nonlinear due to the appearance of $k$ at the denominator, and consequently it makes $P_1$ to be an MINLP problem~\cite{solver}. Thus, in the following section, we first propose an effective method to convert $P_1$ into an equivalent MILP problem, namely $P_2$, by using a linearization technique. After that, $P_2$ can be effectively solved by using commercial solvers. Note that many commercial solvers can find the optimal solution for MILP problems. Moreover, MILP solvers usually can obtain the optimal solution much faster than those of the MINLP solvers~\cite{solver}. As a result, it is very effective to deploy at the server to quickly find the optimal decisions for CDC processes. 
	\subsection{Proposed Linearization Approach}
	In order to transform $P_1$ to an equivalent MILP problem, we first exploit the discrete nature of variable $k$ by introducing the following constraints:
	\begin{align}
		\sum_{j \in \mathcal{N}}jy_{j} = k,  &\label{c11}   \\
	    \sum_{j \in \mathcal{N}}y_{j}=1,&\label{c12}\\
		y_{j} \in \{0,1\},  & \forall j \in \mathcal{N}.   \label{c15}
	\end{align}
	
	Constraints~\eqref{c11} and~\eqref{c12} ensure that the newly introduced binary variables $y_j$ equal 1 only when $j=k$. For example, if $k=2$ then $y_2=1$, while $y_j=0, \forall j\neq k$. 
	Then, constraint~\eqref{c5} can be transformed to a linear format, i.e.,
	\begin{equation}
t_{i} + (1-x_{i})M \geq \dfrac{2\tau_i}{p_i} +\sum_{j=1}^{N} y_j(\dfrac{D}{j\eta_i}+\dfrac{D}{j\eta_i\alpha_i}).\label{c16}
	\end{equation}
$P_2$ can now be defined by objective function~\eqref{c1} and constraints~\eqref{c1}-\eqref{c4} and~\eqref{c6}-\eqref{c16}. Then, we prove in Proposition 2 that $P_2$ is equivalent $P_1$.
	\begin{proposition}
	$P_2$ is equivalent $P_1$.
\end{proposition} 
\begin{IEEEproof}
Without loss of generality, assume $k'$ is the value of $k$ in a feasible solution of the considered MINLP model. Then,~\eqref{c11} becomes 
$\sum_{j \in \mathcal{N}}jy_{j} = k'$. Moreover, since $\sum_{j=1}^{N}y_{j}=1$ (from~\eqref{c12}), we have $y_{k'}=1$ and $y_j=0, \forall j \neq k^*$. Thus, \eqref{c16} becomes:
	\begin{equation}
		\begin{split}
	t_{i} + (1-x_{i})M \geq \dfrac{2\tau_i}{p_i} +\sum_{j \in \mathcal{N}} y_j\bigg(\dfrac{D}{j\eta_i}+\dfrac{D}{j\eta_i\alpha_i}\bigg)\\=\dfrac{2\tau_i}{p_i} +\bigg(\dfrac{D}{k'\eta_i}+\dfrac{D}{k'\eta_i\alpha_i}\bigg),
		\end{split}
\end{equation}
which is equal to~\eqref{c5}. As a result, constraints~\eqref{c11}-\eqref{c16} are equivalent to constraint~\eqref{c5}, and the proof is now completed.
\end{IEEEproof}
As a result of the proposed linearization, $P_2$ can now be effectively solved by commercially available MILP solvers, e.g., CPLEX~\cite{solver}, which can guarantee to find the optimal solution and require much lower solving time compared to that of MINLP solvers.  
\section{Performance Evaluation}
\subsection{Experimental Setup}
We evaluate the performance of the proposed approach in a CDC-based wireless edge network consisting of $N=50$ nodes and 10 tasks with different sizes. Since there is no publicly available dataset for CDC, we adopt two datasets from closely related fields. Particularly, we adopt the task sizes $D$ from the Google Cloud Jobs dataset~\cite{gocj}. Moreover, we arrange the tasks in ascending order of $D$ to clearly show the effects of task sizes on the optimal solutions. Furthermore, we adopt the node capabilities metrics, including $\tau_i$ and $\eta_i$, from the GWA-T traces dataset~\cite{gwat} that contains the performance metrics of virtual machines. For the remaining parameters, we set $p_i=0.9$ and $\alpha_i=2, \forall i \in \mathcal{N}$~\cite{jsac}. Additionally, we compare the proposed approach with the following baseline methods:
\begin{itemize}
	\item \textit{Myopic}: $k$ is set to the maximum possible value, i.e., $k=N$. This is the optimal approach if there is no heterogeneity in the system.
	\item \textit{OneNode}: The server selects only the fastest node. This is equivalent to uncoded and non-distributed computing.
	\item \textit{Static optimal code}~\cite{r1}: $k$ is determined by:
	\begin{equation}
		k= \bigg(1+\dfrac{1}{W_{-1}(-e^{-\overline{\lambda}-1})}\bigg),
	\end{equation}
	where $W_{-1}(.)$ is the lower branch of the Lambert $W$ function and $\overline{\lambda}$ is the average straggling parameter. Since~\cite{r1} does not optimize node selection, we use our node selection optimization to find the optimal results.
\end{itemize}   
\subsection{Simulation Results}
\begin{figure}[!]
	\includegraphics[width=.45\textwidth]{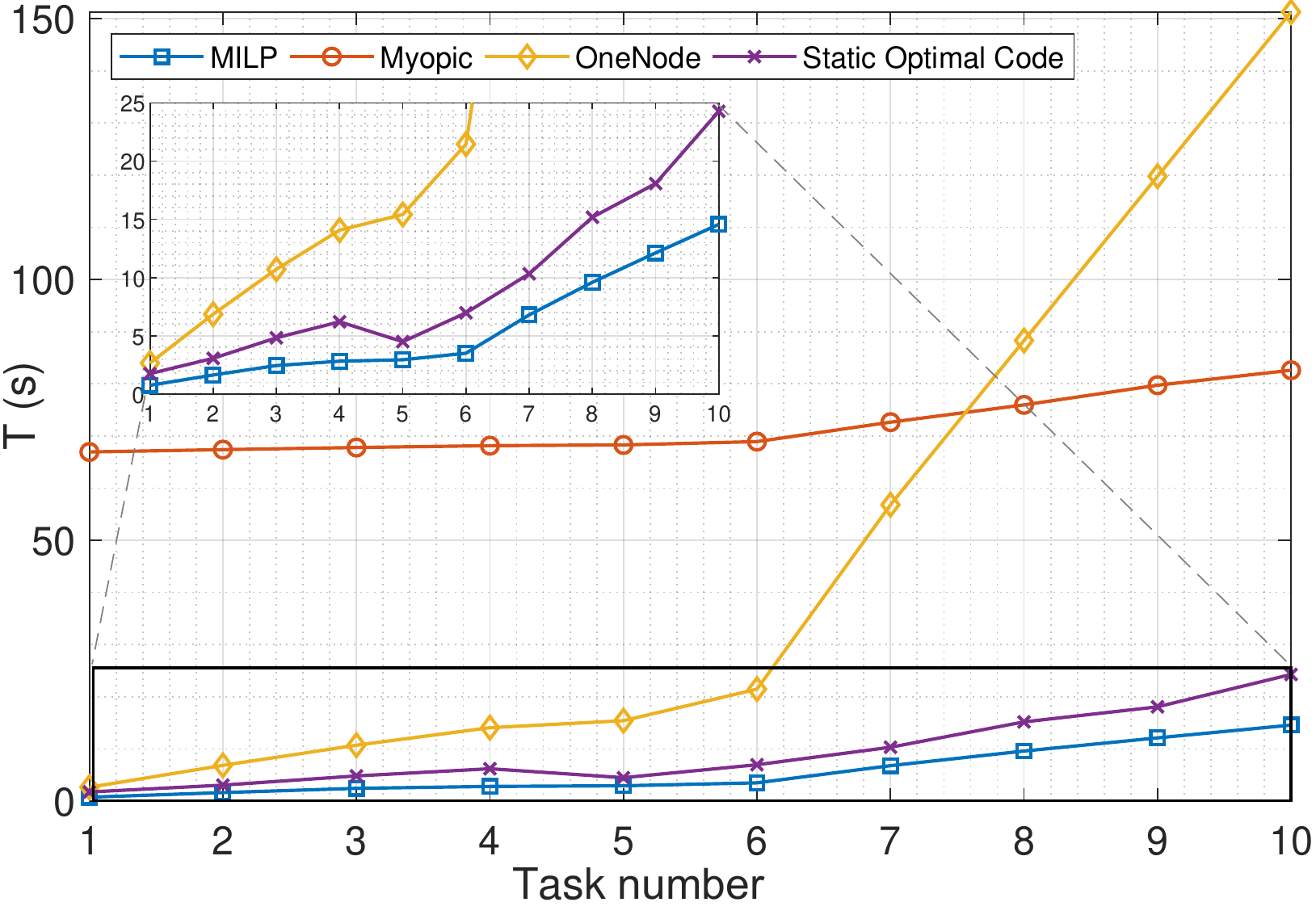}
	\centering
	\caption{Total processing time of the system.}
	\label{Fig:time}
\end{figure}

 Fig.~\ref{Fig:time} shows the total processing time of each task when using different approaches. As illustrated in the figure, our proposed approach outperforms all other approaches for all tasks. Particularly, our proposed approach can achieve $T$ that is up to 90 and 10 times lower than those of the \textit{Myopic} and \textit{OneNode} approaches, respectively. For the static optimal code approach, although it applies the optimal code in~\cite{r1} and our proposed optimization to find the best nodes to perform, its performance is still not as good as that of our proposed framework. Specifically, for the largest-sized task, our proposed solution can reduce the total processing time up to 2.3 times compared with that of the static optimal code solution. The main reason is that, compared to our proposed approach, the static optimal code chooses a higher $k$. Although this reduces the sub-task size, the server has to wait for more nodes to send their results, and thus it may suffer more from the straggling nodes and communication links. 
 
\begin{figure}[!]
	\includegraphics[width=.35\textwidth]{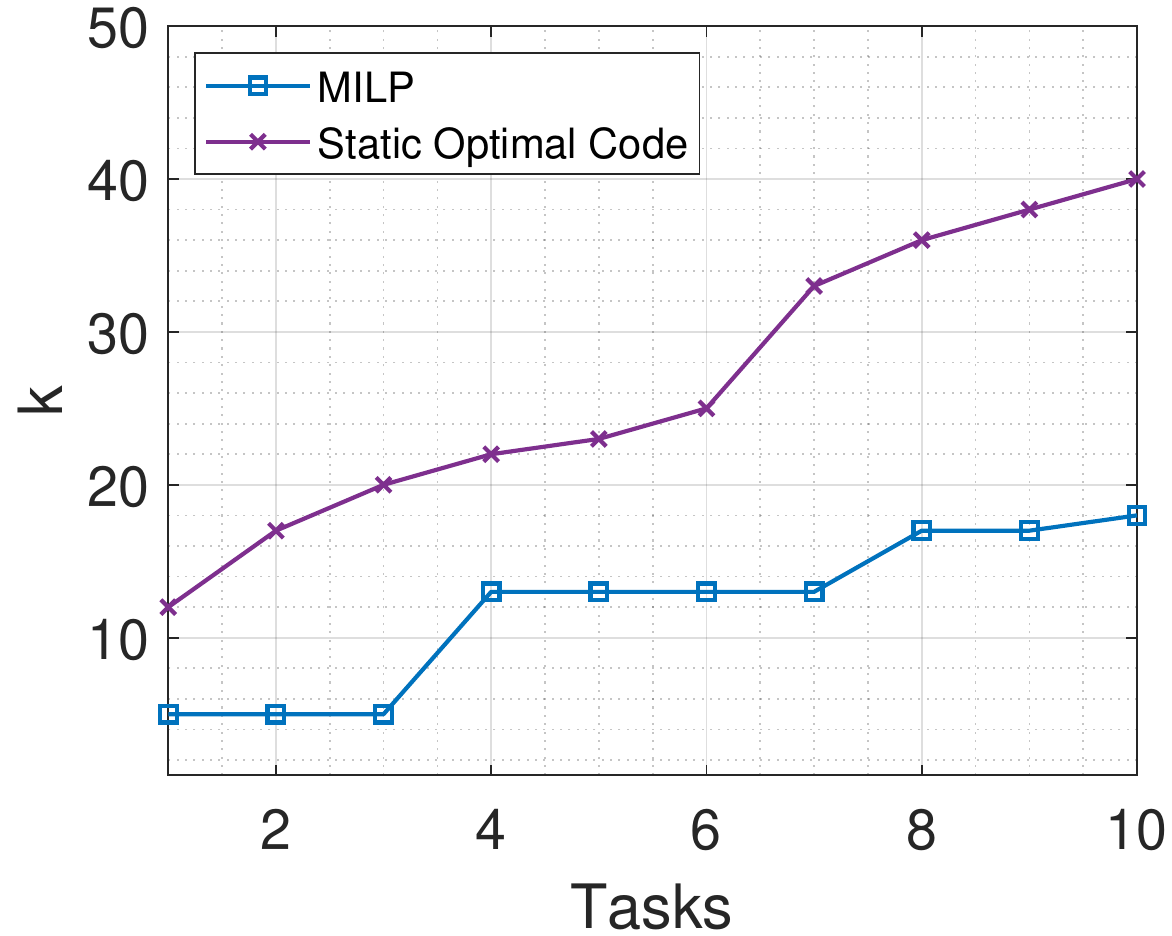}
	\centering
	\caption{The values of $k$.}
	\label{Fig:k}
\end{figure}
 Fig.~\ref{Fig:k} shows the values of $k$ in the solutions obtained by the MILP and the static optimal code. Although their values are different, we can observe that as the task size increases, $k$ also increases. The reason is that, for tasks with smaller sizes, the nodes' communication time has more impacts on the total processing time. In this case, if $k$ is high, the master node has to wait for more nodes. In contrast, when the task size is larger, the nodes' communication time becomes insignificant compared to the computation time. As a result, a higher $k$ is more desirable to reduce the workload at each node, thereby improving the total computation time. However, if $k$ is too high, the delay will be increased due to impacts of the straggling nodes and slow communication links, as observed from the total processing time achieved by our proposed approach and the static optimal code approach in Fig.~\ref{Fig:time}.

\begin{figure}[!]
 	\includegraphics[width=.39\textwidth]{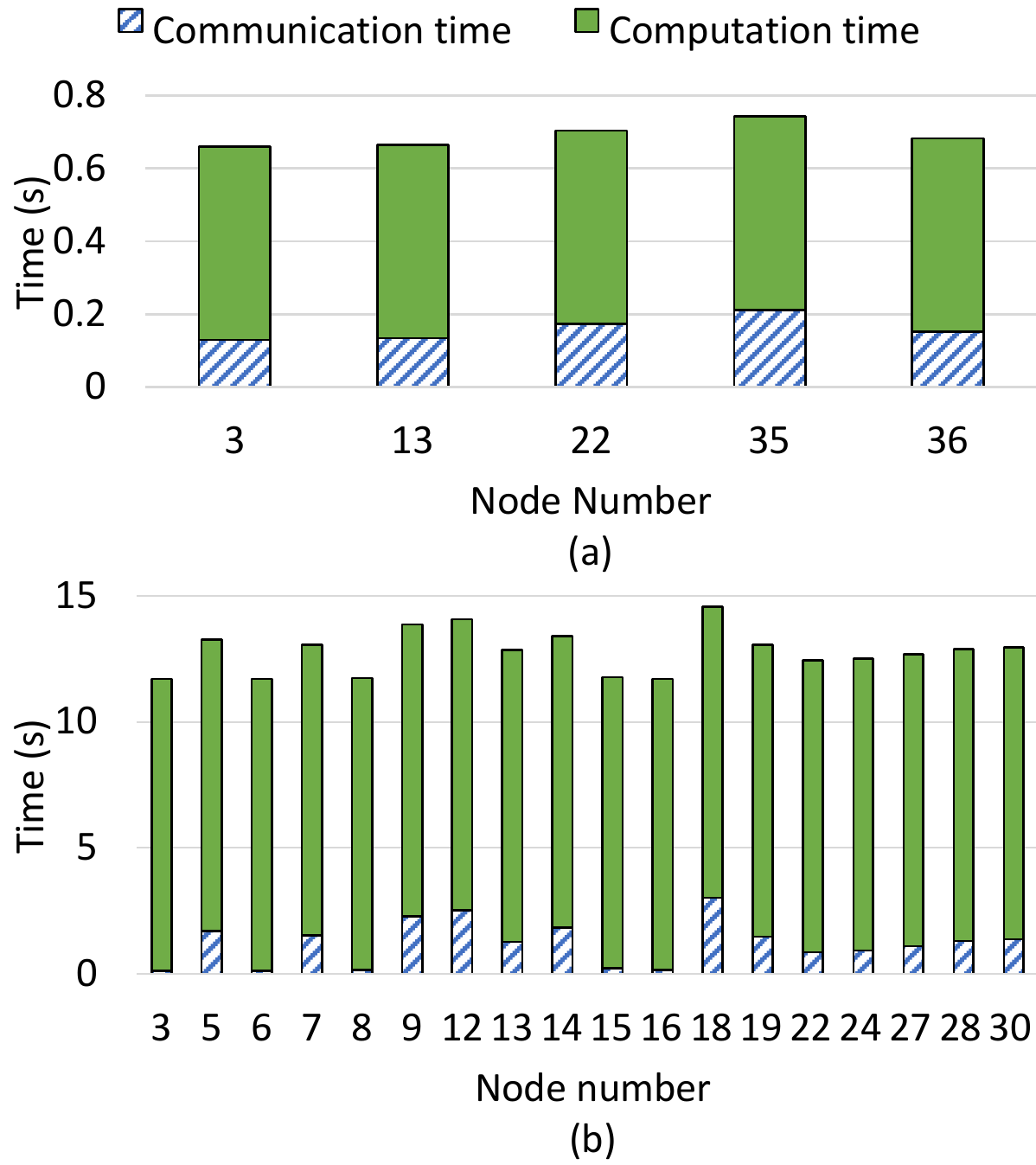}
 	\centering
 	\caption{Processing time components of (a) task 1 and (b) task 10.}
 	\label{Fig:compo_both}
 \end{figure}

Furthermore, to clearly show the relation between the task size and the communication time, we examine the processing time of the nodes in the smallest-size task (task 1) and the largest-sized task (task 10). Note that the optimal $k$ for task 1 is 5, whereas the optimal $k$ for task 10 is 18. As shown in Fig.~\ref{Fig:compo_both}, the communication time is a significant factor in executing task 1. For example, it constitutes up to 19.5\% of the total processing time of node 3 at task 1. In contrast, for task 10, the communication time occupies only 1.1\% of the processing time of the same node. This is also the reason why the optimal values of $k$ for higher in large-sized tasks. Thus, for tasks with large sizes, a high $k$ can significantly reduce the computation time which has much higher impacts on the total processing time.


	\section{Conclusion}
	\label{sec:Sum}
	This letter aims to develop a highly-effective approach to jointly optimize the MDS code and node selection, thereby significantly enhancing the efficiency of CDC in distributed computing over wireless edge networks. Particularly, we have first modeled a joint coding and node selection optimization problem to minimize the processing time for CDC. We have then developed a linearization approach to quickly find the optimal solutions. Simulations based on real-world datasets have been then conducted. The results have shown that our proposed approach can outperform state-of-the-art approach by up to 2.3 times.

\end{document}